\documentclass[letterpaper, 10pt, conference]{ieeeconf}  

\usepackage{graphicx}
\usepackage{pgfplots}
\usepackage{subcaption}
\usepackage{cite}
\usepackage{amsmath,amssymb,amsfonts}
\usepackage{mathrsfs}
\usepackage{algorithmicx}
\usepackage{textcomp}
\usepackage{xcolor}
\usepackage{booktabs}
\usepackage{array}
\usepackage[ruled,lined]{algorithm2e}
\usepackage{multirow}
\usepackage{upgreek}
\usepackage{bm}
\usepackage{bbold}
\usepackage{tikz}
\usepackage{tikz-network}
\usepackage{todonotes}
\usepackage{float} 
\newtheorem{assumption}{Assumption}
\newtheorem{theorem}{Theorem}
\newtheorem{problem}{Problem}
\newtheorem{proposition}{Proposition}
\newtheorem{remark}{Remark}
\newtheorem{corollary}{Corollary}
\newtheorem{definition}{Definition}

\newtheorem{lemma}{Lemma}

\makeatletter
\let\ieeeconf@opargbegintheorem\@opargbegintheorem
\def\@opargbegintheorem#1#2#3{%
  \setbox\@tempboxa\hbox{\hskip\parindent
    \rmfamily\itshape #1\ #2\ (#3):\hskip\labelsep}%
  \ifdim\wd\@tempboxa>\linewidth
    \@IEEEtmpitemindent\itemindent
    \topsep 0pt\rmfamily\trivlist
    \item[]\itemindent\@IEEEtmpitemindent
    \leavevmode\hskip\parindent
    {\itshape #1\ #2\ (#3):}\hskip\labelsep
  \else
    \ieeeconf@opargbegintheorem{#1}{#2}{#3}%
  \fi
  \ignorespaces}
\makeatother

\DeclareMathOperator{\diag}{diag}

\DeclareMathOperator{\dist}{dist}
\DeclareMathOperator{\rank}{rank}
\DeclareMathOperator{\range}{range}
\DeclareMathOperator{\blkdiag}{blkdiag}

\IEEEoverridecommandlockouts                              

\title{\LARGE \bf
Transversal Passivity and Equilibrium Forces in Formation Control}

\author{Feng-Yu Yue and Daniel Zelazo
\thanks{This work was supported by the Israel Science Foundation grant no.\,453/24. Feng-Yu Yue and Daniel Zelazo are with the Stephen B. Klein Faculty of Aerospace Engineering, Technion
– Israel Institute of Technology, Haifa 3200003, Israel. Emails: \textsl{fengyu.yue@campus.technion.ac.il}, \textsl{dzelazo@technion.ac.il.}%
}
}

\begin{document}

\maketitle
\thispagestyle{empty}
\pagestyle{empty}

\begin{abstract}
    We consider controller synthesis for distance-constrained formations, aiming to stabilize the target formation manifold while sustaining admissible equilibrium edge forces. Using the newly introduced notion of transversal passivity, we encode the target formation shape, the prescribed edge lengths, and a subspace of internally balanced edge forces as the port target manifolds of the agents and the edge controller. The synthesis task is thereby reformulated as the joint design of a transversally passive edge controller and its output-target subspace. We then propose a general constructive realization of output-strictly transversally passive systems and specialize it to derive a dynamic edge controller that sustains admissible equilibrium edge forces. Through a compatibility analysis across the interconnection, we establish a closed-loop dissipation margin in terms of the controller parameters and prove local convergence of all agent and controller ports to their target manifolds. Numerical simulations validate the results.
\end{abstract}


\section{Introduction}
Distance-constrained formation control has received sustained attention in cooperative control, motivated by broad applications in mobile-robot teams and multi-vehicle coordination \cite{anderson2008rigid,oh2015survey}. For a group of agents, the objective is to design feedback controllers using relative measurements to achieve prescribed inter-agent distances \cite{oh2015survey,babazadeh2019distance}. Under infinitesimal rigidity, these distance constraints determine the formation shape up to translations and proper rotations, so the admissible target configurations locally form a smooth manifold \cite{asimow1979rigidity,krick2009distance}. Consequently, the control task is naturally formulated as the stabilization of a target formation manifold.

Controller synthesis for this problem must jointly account for the network topology, the agent and controller dynamics, and the nonlinearities induced by the geometric distance constraints. Passivity provides a powerful modular framework for this task by decoupling the component dynamics from the network interconnection, thereby enabling constructive edge-controller synthesis~\cite{arcak2007passivitydesign, Sharf2017a_J}. Existing passivity-based approaches accommodate formation geometry by steering relative edge outputs to prescribed compact target sets~\cite{arcak2007passivitydesign} or by shaping distance-error potentials~\cite{li2025port}. Beyond passivity, edge-space formulations reduce local stability to spectral properties of a linearized edge operator and enable synthesis of stabilizing edge gains via semidefinite programming~\cite{theran2026formation}. 

However, in all of these error-driven constructions, the equilibrium edge forces are not treated as a design freedom: the edge dynamics require every edge force to vanish once the distance errors reach zero. In fact, edge forces need not vanish individually at equilibrium, since internally balanced edge forces (corresponding to \emph{self-stresses}) can remain nonzero without disturbing the agents. Sustaining such internal forces plays a key role in enhancing structural stiffness~\cite{connelly1996stress} and maintaining internal tension~\cite{tognon2018aerial}. This motivates developing a passivity-based edge-controller synthesis where the admissible space of equilibrium edge forces can be explicitly prescribed.

To address this synthesis problem, our recently introduced notion of \emph{transversal passivity}~\cite{yue2026Tpassivity} provides a natural framework. It formulates dissipation relative to prescribed input and output target manifolds while retaining the modularity of classical passivity in decoupling network topology from system dynamics. In the formation-control interconnection, this allows us to encode the rigid-motion orbit of the target shape as the agent output manifold and the prescribed edge lengths as the controller input manifold, while specifying a subspace of internally balanced edge forces as the controller output target. Consequently, the control task is reformulated as the joint design of a transversally passive edge controller and its output-target subspace to achieve shape convergence while sustaining desired equilibrium edge forces.

In this paper, we study the distance-constrained formation controller synthesis problem while sustaining admissible equilibrium edge forces. Our main contributions are as follows. First, we reformulate the considered formation control task as the joint design of a transversally passive edge controller and its output-target subspace. Second, for a smooth input target manifold and a linear output target subspace, we propose a constructive realization of output-strictly transversally passive systems, which specializes to a dynamic edge controller parameterized by the output-target subspace to sustain the corresponding equilibrium forces. Third, through a compatibility analysis under transversal passivity, we establish a closed-loop dissipation margin in terms of the controller parameters and prove local convergence of all agent and controller ports to their target manifolds, thereby achieving the desired formation with equilibrium forces in the prescribed subspace. Although sustaining nonzero equilibrium forces generally leads to a centralized implementation, our numerical simulations validate the theoretical results and demonstrate how a localized subspace choice can preserve these forces with sparser subgraph-confined coupling.

The remainder of this paper is organized as follows. Section~\ref{sec:preliminaries} reviews transversal passivity and distance-constrained formation control. Section~\ref{sec:formation_targets} reformulates the formation control problem within the transversal passivity framework, followed by the constructive controller synthesis and local convergence analysis in Section~\ref{sec:synthesis_convergence}. Numerical simulations and concluding remarks are given in Sections~\ref{sec:numerical_results} and~\ref{sec:conclusion}, respectively.

\paragraph*{Notations}
All inner products $\langle\cdot,\cdot\rangle$ and norms $\|\cdot\|$ are Euclidean. We denote by $I_n$ the $n\times n$ identity matrix, by $\mathbf{1}_n\in\mathbb{R}^n$ the all-ones vector, by $\otimes$ the Kronecker product, and by $SO(2)$ the group of $2\times 2$ proper rotation matrices. 

For a set $\mathcal{S}$, $|\mathcal{S}|$ denotes its cardinality, and $\dist(x,\mathcal{S}):=\inf_{y\in\mathcal{S}}\|x-y\|$ denotes the Euclidean distance from $x\in\mathbb{R}^n$ to a nonempty $\mathcal{S}\subseteq\mathbb{R}^n$. For a linear subspace $\mathcal{L}\subseteq\mathbb{R}^n$, $\mathrm{P}_{\mathcal{L}}$ and $\mathrm{P}_{\mathcal{L}^\perp}=I_n-\mathrm{P}_{\mathcal{L}}$ denote the orthogonal projectors onto $\mathcal{L}$ and its orthogonal complement $\mathcal{L}^\perp$, respectively. For a smooth manifold $\mathcal{M}\subseteq\mathbb{R}^n$, $T_p\mathcal{M}$ and $N_p\mathcal{M}$ denote its tangent and normal spaces at $p\in\mathcal{M}$, respectively.

An undirected graph $\mathcal{G}=(\mathcal{V},\mathcal{E})$ consists of a finite vertex set $\mathcal{V}$ and an edge set $\mathcal{E}\subseteq\mathcal{V}\times\mathcal{V}$. Assigning an arbitrary orientation to each edge $e_k\in\mathcal{E}$, the incidence matrix $E\in\mathbb{R}^{|\mathcal{V}|\times|\mathcal{E}|}$ is defined entrywise by $[E]_{ik}:=1$ if vertex $i$ is the head of $e_k$, $[E]_{ik}:=-1$ if $i$ is the tail of $e_k$, and $[E]_{ik}:=0$ otherwise.

\section{Preliminaries}\label{sec:preliminaries}
This section introduces necessary background on transversal passivity and the distance-constrained formation control.
\subsection{Transversal Passivity}\label{subsec:transversal_passivity}
Consider two nonlinear systems
$$   \Sigma_i:\ \ \dot x_i = f_i(x_i,u_i), \quad y_i = h_i(x_i,u_i), \qquad i=1,2, $$
with state $x_i\in\mathcal X_i\subseteq\mathbb R^{n_i}$, input $u_i\in\mathcal U_i\subseteq\mathbb R^{m_i}$, and output $y_i\in\mathcal Y_i\subseteq\mathbb R^{m_i}$, where $\mathcal X_i,\mathcal U_i,\mathcal Y_i$ are open, $f_i$ is locally Lipschitz, and $h_i$ is $C^1$. The two systems are interconnected via the power-conserving linear feedback law
\begin{equation}\label{eq:general_interconnection}
  u_1 = -\Gamma^\top y_2, \qquad u_2 = \Gamma y_1,
\end{equation}
where $\Gamma\in\mathbb R^{m_2\times m_1}$ is the \emph{forward channel} and $-\Gamma^\top$ is the \emph{reverse channel}. 

Equip each $\Sigma_i$ with embedded target submanifolds $\mathcal M_{u_i}\subseteq\mathcal U_i$ and $\mathcal M_{y_i}\subseteq\mathcal Y_i$. On tubular neighborhoods $\mathcal N_{u_i}\subseteq\mathcal U_i$ and $\mathcal N_{y_i}\subseteq\mathcal Y_i$ where the Euclidean nearest-point projections $r_{u_i}:\mathcal N_{u_i}\to\mathcal M_{u_i}$ and $r_{y_i}:\mathcal N_{y_i}\to\mathcal M_{y_i}$ are uniquely defined and smooth, the \emph{transverse port error maps} are defined as
\begin{equation}\label{eq:port_errors}
  e_{u_i}(u_i) := u_i - r_{u_i}(u_i), \qquad e_{y_i}(y_i) := y_i - r_{y_i}(y_i),
\end{equation}
These error maps represent the normal displacements from the targets and satisfy $\|e_{u_i}(u_i)\|=\dist(u_i,\mathcal M_{u_i})$ and $\|e_{y_i}(y_i)\|=\dist(y_i,\mathcal M_{y_i})$. Define the tubular domain $\mathcal D_i:=\{(x_i,u_i)\in\mathcal X_i\times\mathcal N_{u_i} : h_i(x_i,u_i)\in\mathcal N_{y_i}\}$.
\begin{definition}[Transversal passivity~\cite{yue2026Tpassivity}]\label{def:transversal_passivity}
The system $\Sigma_i$ is called \emph{$(\mathcal M_{u_i},\mathcal M_{y_i})$-passive} on $\mathcal D_i\ne\emptyset$ if there exist a continuously differentiable storage function $V_i:\mathcal X_i\to\mathbb R_{\ge 0}$ and constants $\varepsilon_i,\gamma_i\ge 0$ such that, for all $(x_i,u_i)\in\mathcal D_i$,
$$
  \dot V_i(x_i,u_i) \le \langle e_{u_i}(u_i), e_{y_i}(y_i)\rangle - \varepsilon_i\|e_{y_i}(y_i)\|^2 - \gamma_i\|e_{u_i}(u_i)\|^2.
$$
$\Sigma_i$ is \emph{output-strictly} $(\mathcal M_{u_i},\mathcal M_{y_i})$-passive if $\varepsilon_i>0$.
\end{definition}

Transversal passivity penalizes only deviations normal to $(\mathcal M_{u_i},\mathcal M_{y_i})$ while leaving motions along the target manifolds unpenalized.

Suppose now that both subsystems $\Sigma_i$ in~\eqref{eq:general_interconnection} are $(\mathcal M_{u_i},\mathcal M_{y_i})$-passive with storage functions $V_i(x_i)$ and constants $\varepsilon_i,\gamma_i\ge 0$. Although~\eqref{eq:general_interconnection} is power-conserving ($\langle u_1,y_1\rangle+\langle u_2,y_2\rangle = 0$), this cancellation generally fails under transversal passivity because the supply rates are evaluated at the transverse port errors rather than the port signals. To quantify this mismatch for the forward channel $\Gamma$, we define the \emph{compatibility defect}~\cite[Definitions~2--3]{yue2026Tpassivity}
$$
  \delta_{\mathrm{fc}}(y_1) := e_{u_2}(\Gamma y_1) - \Gamma\,e_{y_1}(y_1) = \Gamma\,r_{y_1}(y_1) - r_{u_2}(\Gamma y_1),
$$
which measures the failure of $\Gamma$ to commute with the nearest-point projections. The forward channel $\Gamma$ is called \emph{exactly compatible} on a set $\mathcal S$ if $\delta_{\mathrm{fc}}(y_1)=0$ for all $y_1\in\mathcal S$, and \emph{first-order compatible} along $\mathcal S$ if $\delta_{\mathrm{fc}}(y_1)=0$ and $\mathrm d_{y_1}\delta_{\mathrm{fc}} = 0$ for all $y_1\in\mathcal S$. The defect $\delta_{\mathrm{rc}}(y_2)$ and compatibility notions for the reverse channel $-\Gamma^\top$ are defined analogously.

For the feedback interconnection~\eqref{eq:general_interconnection} with composite storage $V_{\mathrm{cl}}(x):=V_1(x_1)+V_2(x_2)$ and $x=(x_1,x_2)$, the closed-loop dissipation satisfies~\cite[Theorem~1]{yue2026Tpassivity} $\dot V_{\mathrm{cl}}(x) \le s_\delta(x) - W_0(x)$, where $W_0(x)$ collects the subsystem dissipation and
$$
  s_\delta(x) := \langle\delta_{\mathrm{rc}}(y_2(x)),e_{y_1}(y_1(x))\rangle + \langle\delta_{\mathrm{fc}}(y_1(x)),e_{y_2}(y_2(x))\rangle
$$
is the \emph{residual transverse supply}. Since $e_{y_i}$, $\delta_{\mathrm{fc}}$, and $\delta_{\mathrm{rc}}$ vanish when all ports lie on their target manifolds, $s_\delta$ is locally at least quadratic, yielding three regimes: (i) \emph{vanishing residual} ($s_\delta\equiv 0$), ensured if both channels are exactly compatible; (ii) \emph{quadratic cancellation} ($s_\delta$ is locally cubic), ensured if both channels are first-order compatible; and (iii) \emph{nonzero quadratic residual}, where $W_0$ must dominate $s_\delta$ at second order for net dissipation. Combined with compact confinement, port convergence guarantees across these three regimes are established in~\cite[Corollary~2, Theorems~2 and~3]{yue2026Tpassivity}.

\subsection{Graph Rigidity and Distance-Constrained Formations}
Consider $n\ge 3$ single-integrator agents in $\mathbb{R}^2$ with dynamics $\dot p_i = u_i$, where $p_i, u_i\in\mathbb{R}^2$ denote the position and control input of agent $i\in\mathcal{V}$. The inter-agent interaction topology is modeled by a connected undirected graph $\mathcal{G}=(\mathcal{V},\mathcal{E})$ with $|\mathcal{E}|=m$ edges. The formation is represented by a \emph{framework} $(\mathcal{G},p)$, where $p=(p_1^\top,\ldots,p_n^\top)^\top\in\mathbb{R}^{2n}$ is the configuration of the agents, yielding $\dot p = u$. Let $\bar E := E\otimes I_2 \in \mathbb{R}^{2n\times 2m}$, so that the $k$-th block $(\bar E^\top p)_k = p_i - p_j$ gives the relative position of edge $e_k=(i,j)\in\mathcal{E}$. 

The shape of $(\mathcal{G},p)$ is described by the \emph{rigidity function}
$$\mathrm{r}_{\mathcal{G}}(p) = \frac12 (\|(\bar E^\top p)_1\|^2, \ldots, \|(\bar E^\top p)_m\|^2)^\top \in \mathbb{R}^m,$$
with
\[
\thickmuskip=3mu\thinmuskip=2mu\relax R_{\mathcal{G}}(p) := \frac{\partial \mathrm{r}_{\mathcal{G}}}{\partial p}(p)=\blkdiag((\bar E^\top p)_1^\top,\ldots,(\bar E^\top p)_m^\top)\bar E^\top
\]
being the \emph{rigidity matrix}. For a framework $(\mathcal{G},p)$, a \emph{stress} is a vector of scalars $\omega=(\omega_1,\ldots,\omega_m)^\top\in\mathbb{R}^m$ assigned to the edges. A stress is called a \emph{self-stress} if the induced edge forces $\bigl(\omega_1(\bar E^\top p)_1^\top,\ldots,\omega_m(\bar E^\top p)_m^\top\bigr)^\top\in\mathbb{R}^{2m}$ are balanced at every node, i.e.,
\begin{equation}\label{eq:self_stress}
\bar E\bigl(\omega_1(\bar E^\top p)_1^\top,\ldots,\omega_m(\bar E^\top p)_m^\top\bigr)^\top = R_{\mathcal{G}}(p)^\top\omega = 0.
\end{equation}
Thus, $\ker(R_{\mathcal{G}}(p)^\top)$ is the space of self-stresses of $(\mathcal{G},p)$~\cite{asimow1978rigidity,connelly2022frameworks}.

Let $p^\star\in\mathbb{R}^{2n}$ be a target configuration with desired edge lengths $d=(d_1,\ldots,d_m)^\top$, where $d_k=\|(\bar E^\top p^\star)_k\|>0$ for all $e_k\in\mathcal{E}$. Throughout this paper, we assume that the target framework $(\mathcal{G}, p^\star)$ is \emph{infinitesimally rigid}, meaning that $\rank(R_{\mathcal{G}}(p^\star)) = 2n - 3$~\cite{asimow1979rigidity,laman1970rigidity}. In this case, $\dim\ker(R_{\mathcal{G}}(p^\star)^\top) = m - (2n-3)$, and the only infinitesimal motions at $p^\star$ that preserve the edge lengths are planar translations and rotations. The control objective is to design the control input $u$ such that the agents locally converge to the target formation matching $p^\star$ up to planar translations and proper rotations, while sustaining desired edge forces at equilibrium.

\section{Problem Reformulation via Transversal Passivity}\label{sec:formation_targets}
This section reformulates the distance-constrained formation control problem within the transversal passivity framework~\cite{yue2026Tpassivity}. Whereas classical passivity measures the supply $\langle u,y\rangle$ relative to the origin to certify convergence to a point equilibrium, the agent positions here must converge to the noncompact manifold of configurations with the desired shape. Transversal passivity accommodates this objective by evaluating the supply $\langle e_u,e_y\rangle$ via the transverse port errors~\eqref{eq:port_errors} relative to prescribed \emph{target manifolds}. The design thus begins by encoding the control objective as agent target manifolds, from which consistent controller targets follow through the interconnection channels.

\subsection{Feedback Interconnection with Edge Ports}
\begin{figure}[t]
\centering
\begin{tikzpicture}[
  >={Latex[length=2.2mm,width=1.6mm]},
  sys/.style={draw, thick, rounded corners=2pt, fill=blue!6,
              minimum width=13mm, minimum height=8.5mm},
  chan/.style={draw, thick, fill=gray!12,
              minimum width=10mm, minimum height=7mm},
  sig/.style={thick, ->, rounded corners=3pt},
  lbl/.style={font=\small, inner sep=2pt},
  tgt/.style={font=\small, inner sep=2pt, text=blue!60!black},
  note/.style={font=\scriptsize, text=gray!70!black, align=center, inner sep=1pt}
]
  \node[sys]  (S1) at (0, 0)      {$\Sigma_1$};
  \node[sys]  (S2) at (0,-2.0)    {$\Sigma_2$};
  \node[chan] (A)  at ( 2.1,-1.0) {$\bar E^\top$};
  \node[chan] (AT) at (-2.1,-1.0) {$-\bar E$};
  \draw[sig] (S1.east) -- node[lbl,above]{$y_1$} node[tgt,below]{$\mathcal M_{y_1}$} (A.north |- S1.east) -- (A.north);
  \draw[sig] (A.south) -- (A.south |- S2.east) -- node[lbl,below]{$u_2$} node[tgt,above]{$\mathcal C$} (S2.east);
  \draw[sig] (S2.west) -- node[lbl,below]{$y_2$} node[tgt,above]{$\mathcal Z$} (AT.south |- S2.west) -- (AT.south);
  \draw[sig] (AT.north) -- (AT.north |- S1.west) -- node[lbl,above]{$u_1$} node[tgt,below]{$\{0\}$} (S1.west);
  \node[note, anchor=west] at ($(A.east)+(0.08,0)$) {forward\\$\bar E^\top\mathcal M_{y_1}\subseteq\mathcal C$};
  \node[note, anchor=east] at ($(AT.west)+(-0.08,0)$) {reverse\\$-\bar E\,\mathcal Z=\{0\}$};
\end{tikzpicture}
\caption{Feedback interconnection of the agents $\Sigma_1$ and the edge controller $\Sigma_2$ through the forward channel $\bar E^\top$ and the reverse channel $-\bar E$. Each port signal is annotated with its target manifold (blue); the channel annotations relate the targets across the interconnection.}
\label{fig:feedback_interconnection}
\end{figure}
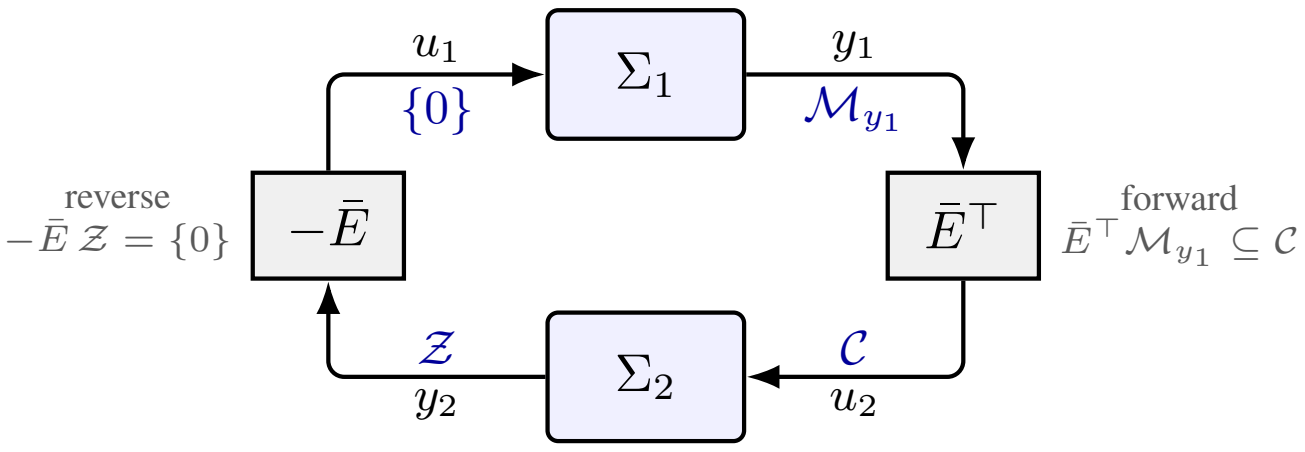

Consider the feedback interconnection in Fig.~\ref{fig:feedback_interconnection}, where the plant $\Sigma_1$ represents the stacked agent dynamics with input $u_1:=u$ and output $y_1:=p$,
\begin{equation}\label{eq:sys1}
  \Sigma_1:\quad \dot{p} = u_1, \qquad y_1 = p,
\end{equation}
and $\Sigma_2$ is the edge controller to be designed. Choosing the forward channel $\bar E^\top$ and reverse channel $-\bar E$ yields the edge relative positions $u_2 = \bar E^\top p$ and plant input $u_1 = -\bar E y_2$.

\subsection{Agent Target Manifolds}
The targets of $\Sigma_1$ are dictated by the control objective rather than chosen. First, at steady state the agents should be at rest, so the input target is
\begin{equation}\label{eq:input_target_plant}
  \mathcal M_{u_1} = \{0\}.
\end{equation}
Second, the agent positions should form the desired shape. Since the edge lengths determine $p^\star$ only up to planar translations and rotations, the output target cannot be a single point. Instead, we take the orbit of $p^\star$ under planar rigid motions,
\begin{equation}\label{eq:formation_target}
  \mathcal M_{y_1} = \bigl\{ (I_n\otimes Q)p^\star+\mathbf 1_n\otimes \eta : Q\in SO(2),\ \eta\in\mathbb R^2 \bigr\}.
\end{equation}
Since $d_k>0$, the only planar rigid motion fixing $p^\star$ is the identity, and $SE(2)$ acts properly on $\mathbb R^{2n}$. Therefore, $\mathcal M_{y_1}$ is a connected, closed, embedded three-dimensional submanifold of $\mathbb R^{2n}$ diffeomorphic to $SE(2)$~\cite{lee2024smoothmanifolds}. Moreover, by infinitesimal rigidity, $\mathcal M_{y_1}$ coincides near $p^\star$ with the set of configurations satisfying all distance constraints, while excluding reflections and other incongruent realizations of the same distances.

Since $\mathcal M_{y_1}$ is a closed embedded submanifold, it admits a tubular neighborhood $\mathcal N_{y_1}\subseteq\mathbb R^{2n}$ on which the nearest-point projection $r_{y_1}$ is well defined and smooth, with transverse error $e_{y_1}(p)=p-r_{y_1}(p)$. On the input side, we take $\mathcal N_{u_1}=\mathbb R^{2n}$, $r_{u_1}\equiv 0$, and $e_{u_1}(u_1)=u_1$. Choosing the output-distance storage function
\begin{equation}\label{eq:plant_storage}
  V_1(p)=\frac12\dist(p,\mathcal M_{y_1})^2=\frac12\|e_{y_1}(p)\|^2
\end{equation}
and differentiating $V_1$ along $\Sigma_1$ yields $\dot V_1 = \langle e_{y_1}(p), u_1 \rangle = \langle e_{u_1}(u_1), e_{y_1}(y_1) \rangle$. Thus, $\Sigma_1$ is $(\{0\}, \mathcal M_{y_1})$-passive on $\mathcal N_{y_1} \times \mathbb R^{2n}$~\cite[Proposition~2]{yue2026Tpassivity}. Note that $r_{y_1}$ is an analysis tool and is not computed by the agents.

\subsection{Controller Target Manifolds}
\begin{figure}[t]
\centering
\begin{tikzpicture}[>={Latex[length=2mm,width=1.4mm]}, font=\small]
  \def\d{1.55}\def\L{2.45}\def\th{30}\def\ph{68}
  \colorlet{tgt}{blue!60!black}
  \colorlet{err}{orange!85!black}
  \fill[tgt!5] (0,0) -- (-12:\d) arc[start angle=-12, end angle=108, radius=\d] -- cycle;
  \draw[tgt, thick] (-12:\d) arc[start angle=-12, end angle=108, radius=\d];
  \node[tgt, anchor=north west, inner sep=1pt] at (-12:\d) {$d_kS^1$};
  \draw[gray!65] ($(\th:\d)+(\th:0.13)$) -- ++(\th-90:0.13) -- ++(\th+180:0.13);
  \draw[gray!70!black, densely dotted, thick] (\th:\d) -- (\ph:\d);
  \fill (0,0) circle (1.3pt) node[below left, inner sep=1.5pt]{$0$};
  \draw[->, thick] (0,0) -- (\th:\L) node[right, inner sep=2pt]{$u_{2,k}$};
  \draw[err, line width=2.2pt, line cap=round] (\th:\d+0.04) -- (\th:\L-0.2);
  \node[err, anchor=north west, inner sep=1pt] at ($(\th:{(\d+\L)/2})+(\th-90:0.08)$) {$\ell_k$};
  \node[anchor=south east, inner sep=1pt] at ($(\th:\d/2)+(\th+90:0.04)$) {$d_k$};
  \fill[tgt] (\th:\d) circle (1.8pt);
  \node[tgt, anchor=north east, inner sep=1pt] at (1.46,0.36) {$r_{u_2,k}$};
  \draw[gray!70!black, fill=white] (\ph:\d) circle (1.8pt);
  \node[gray!70!black, anchor=south west, inner sep=1pt] at ($(\ph:\d)+(0.05,0.05)$) {$(\bar E^\top r_{y_1}(p))_k$};
\end{tikzpicture}
\caption{Edgewise projection onto $\mathcal C$ for edge $k$. The nearest point $r_{u_2,k}=d_k\widehat u_{2,k}$ on the target circle $d_kS^1$ (blue) is obtained by rescaling $u_{2,k}$, and the transverse error $e_{u_2,k}=\ell_k\widehat u_{2,k}$ (orange) is normal to the circle. The hollow point is the $k$-th edge of the nearest target formation $r_{y_1}(p)$; the dotted chord to it is the forward compatibility defect analyzed in Section~\ref{sec:synthesis_convergence}.}
\label{fig:edge_projection}
\end{figure}
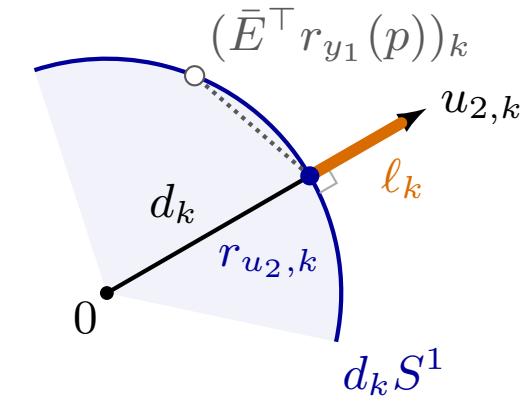

In contrast to the agent targets, the controller targets are design choices, constrained only by consistency with the agent targets through the interconnection channels. At the input port, each edge attains its desired length $d_k$ if and only if $u_{2,k} \in d_k S^1 := \{x \in \mathbb{R}^2 : \|x\| = d_k\}$. At the output port, since $u_1 = -\bar E y_2$, the constraint $\mathcal{M}_{u_1} = \{0\}$ requires the target outputs to lie in $\ker \bar E$, which by~\eqref{eq:self_stress} also contains the edge forces induced by any self-stress $\omega\in\ker(R_{\mathcal{G}}(p)^\top)$. Accordingly, we choose the controller target manifolds as
\begin{equation}\label{eq:input_target}
  \mathcal M_{u_2} = \mathcal C := \prod_{k=1}^{m} d_k S^1, \qquad
  \mathcal M_{y_2} = \mathcal Z \subseteq \ker \bar E.
\end{equation}
We choose $\mathcal C$ rather than the realizable image $\bar E^\top\mathcal M_{y_1}=\{(I_m\otimes Q)\bar E^\top p^\star : Q\in SO(2)\}$ as the controller input target because its nearest-point projection is edgewise: projecting onto $\mathcal C$ only rescales each $u_{2,k}$ to length $d_k$ (Fig.~\ref{fig:edge_projection}), whereas projecting onto $\bar E^\top\mathcal M_{y_1}$ requires fitting a common rotation to all edges simultaneously. This relaxation introduces no spurious targets: since infinitesimal rigidity implies local rigidity~\cite{asimow1978rigidity}, $\mathcal C\cap\range\bar E^\top=\bar E^\top\mathcal M_{y_1}$ in a neighborhood of $\bar E^\top\mathcal M_{y_1}$. Meanwhile, the linear subspace $\mathcal Z \subseteq \ker \bar E$ is selected to prescribe the admissible balanced edge forces at equilibrium.

For $\Sigma_2$, the edgewise projection onto $\mathcal C$ is smooth on $\mathcal N_{u_2} := \{u_2 \in \mathbb R^{2m} : u_{2,k} \ne 0,\ \forall k \in \{1,\ldots,m\}\}$, while for the linear subspace $\mathcal Z\subseteq\ker \bar E$ the nearest-point projection is the orthogonal projector $\mathrm P_{\mathcal Z}$, which is defined globally, i.e., $\mathcal N_{y_2}=\mathbb R^{2m}$. On $\mathcal N_{u_2}$, define the edge-length error $\ell_k := \|u_{2,k}\| - d_k$ and unit direction $\widehat u_{2,k} := u_{2,k}/\|u_{2,k}\|$, and stack them as $\ell := (\ell_1,\ldots,\ell_m)^\top$, and
\begin{equation}\label{eq:unit_direction_matrix}
U(u_2) := \operatorname{blkdiag}(\widehat u_{2,1},\ldots,\widehat u_{2,m}),
\end{equation}
which satisfies $U(u_2)^\top U(u_2) = I_m$. The port projections and transverse errors of $\Sigma_2$ are then given by
\begin{equation}\label{eq:sigma2_targetmaps}
\begin{aligned}
  r_{u_2}(u_2) &= U(u_2)d, & e_{u_2}(u_2) &= U(u_2)\ell,\\
  r_{y_2}(y_2) &= \mathrm P_{\mathcal Z}\,y_2, & e_{y_2}(y_2) &= \mathrm P_{\mathcal Z^\perp}\,y_2,
\end{aligned}
\end{equation}
with $\|e_{u_2}(u_2)\| = \|\ell\|$.


\subsection{Problem Formulation}
At this point, the agent side of the interconnection is fully specified: $\Sigma_1$ is $(\{0\},\mathcal M_{y_1})$-passive, and the controller input target is fixed as $\mathcal M_{u_2}=\mathcal C$. What remains is the edge controller $\Sigma_2$ together with its output target $\mathcal Z$, which must be chosen jointly. The formation control task thus reduces to the following synthesis problem.

\begin{problem}\label{prob:synthesis}
Select a linear output-target subspace $\mathcal M_{y_2} = \mathcal Z \subseteq \ker \bar E$ and synthesize a dynamic edge controller $\Sigma_2$ such that:
\begin{enumerate}
    \item[i)] $\Sigma_2$ is output-strictly $(\mathcal C, \mathcal Z)$-passive;
    \item[ii)] at any target configuration $p\in\mathcal M_{y_1}$, the closed-loop equilibria sustain the balanced edge forces $y_2\in\mathcal Z$ induced by the corresponding self-stresses in $\ker(R_{\mathcal G}(p)^\top)$; and
    \item[iii)] the closed-loop system achieves local convergence, with all realized port signals $(u_1,y_1,u_2,y_2)$ converging to $(\mathcal M_{u_1},\mathcal M_{y_1},\mathcal C,\mathcal Z)$.
\end{enumerate}
\end{problem}

Two features distinguish Problem~\ref{prob:synthesis} from a standard passivity-based interconnection. First, since $\mathcal C$ relaxes $\bar E^\top\mathcal M_{y_1}$, projecting onto the targets does not commute with the forward channel, i.e., the interconnection carries a nonzero compatibility defect, so the general interconnection results of~\cite{yue2026Tpassivity} do not apply directly. Second, the choice of $\mathcal Z$ governs both the equilibrium edge forces sustained and the inter-edge coupling of the controller, with the two extremes $\mathcal Z=\{0\}$ and $\mathcal Z=\ker\bar E$ trading one against the other.

Section~\ref{sec:synthesis_convergence} addresses Problem~\ref{prob:synthesis}: for a given $\mathcal Z$, it constructs an output-strictly transversally passive edge controller (Proposition~\ref{prop:controller_certificate}), shows that the compatibility defect enters the closed-loop energy balance only at second order (Proposition~\ref{prop:channel_compatibility} and Lemma~\ref{lem:net_dissipation}), establishes local convergence (Theorem~\ref{thm:local_convergence}), and examines how the choice of $\mathcal Z$ trades off retained equilibrium forces against inter-edge coupling. Section~\ref{sec:numerical_results} then compares decoupled, centralized, and localized choices of $\mathcal Z$ numerically.

\section{Constructive Controller Synthesis and Local Convergence}\label{sec:synthesis_convergence}
In this section, we address Problem~\ref{prob:synthesis} for a given subspace $\mathcal Z\subseteq \ker \bar E$. We synthesize an output-strictly $(\mathcal M_{u_2},\mathcal M_{y_2})$-passive dynamical controller $\Sigma_2$ and establish local closed-loop port convergence.
For brevity, the arguments of $e_{u_2}(u_2)$, $e_{y_2}(y_2)$ and $U(u_2)$ are omitted when they are clear from the context.

\subsection{Construction from the Transverse Supply}
We first present a general construction of output-strictly transversally passive systems for an arbitrary smooth input target manifold $\mathcal M_u$ (with transverse error $e_u$ on $\mathcal N_u$) and a linear output target subspace $\mathcal M_y$ (with $e_y=\mathrm P_{\mathcal M_y^\perp}y$). 

Starting from the transverse supply $\langle e_u,e_y\rangle$, we seek state dynamics and an output map satisfying the output-strict dissipation inequality
\[
\dot V \le \langle e_u,e_y\rangle - \varepsilon\|e_y\|^2,
\]
with $\varepsilon>0$. For a state-only storage function $V(\xi)$, we have $\dot V=\langle\nabla V(\xi),\dot\xi\rangle$, whereas the linearity of $\mathcal M_y$ gives $\langle e_u,e_y\rangle = \langle \mathrm P_{\mathcal M_y^\perp}e_u,y\rangle$. This suggests the classical port-Hamiltonian structure~\cite{vanderschaft2014port,vanderschaft2017L2}: taking $V$ as the Hamiltonian, we introduce a matrix-valued design map $L(u)$ from the state space to the output space, drive $\xi$ by the projected transverse input error through $L(u)^\top$, and take the collocated output $L(u)\nabla V(\xi)$, complemented by damping and a direct feedthrough of $e_u$. This leads to the following result.

\begin{proposition}[Constructive transversal passivity realization]\label{prop:controller_certificate}
Let $V:\mathcal X\to\mathbb R_{\ge 0}$ be $C^2$ on an open state domain $\mathcal X\subseteq\mathbb R^{n_\xi}$, and let $L(u)$ be a $C^1$ matrix-valued map of compatible dimensions on $\mathcal N_u$. For any gains $a,b,g>0$ and any linear subspace $\mathcal M_y$, the system
\begin{equation}\label{eq:general_tp_controller}
\begin{cases}
\dot\xi = L(u)^\top \mathrm P_{\mathcal M_y^\perp}\bigl(b\,e_u - a\,L(u)\nabla V(\xi)\bigr),\\
y = b\,L(u)\nabla V(\xi) + g\,e_u,
\end{cases}
\end{equation}
is output-strictly $(\mathcal M_u,\mathcal M_y)$-passive on $\mathcal X\times\mathcal N_u$ with storage function $V(\xi)$, satisfying
\begin{equation}\label{eq:controller_storage_identity}
\begin{aligned}
\dot V &= \langle e_u,e_y\rangle - a\|\mathrm P_{\mathcal M_y^\perp}L\nabla V\|^2 - g\|\mathrm P_{\mathcal M_y^\perp}e_u\|^2 \\
&\le \langle e_u,e_y\rangle - \frac{a}{b^2+ag}\|e_y\|^2.
\end{aligned}
\end{equation}
Moreover, at any held target input $u\in\mathcal M_u$, a state $\xi\in\mathcal X$ is an equilibrium of~\eqref{eq:general_tp_controller} if and only if $L(u)\nabla V(\xi)\in\mathcal M_y$. 
\end{proposition}

\begin{proof}
By~\eqref{eq:general_tp_controller}, $e_y=\mathrm P_{\mathcal M_y^\perp}y = b\,\mathrm P_{\mathcal M_y^\perp}L\nabla V + g\,\mathrm P_{\mathcal M_y^\perp}e_u$. Differentiating $V$ along~\eqref{eq:general_tp_controller} and substituting $b\,\mathrm P_{\mathcal M_y^\perp}L\nabla V = e_y - g\,\mathrm P_{\mathcal M_y^\perp}e_u$ into $\dot V = \nabla V^\top\dot\xi$ yields the equality in~\eqref{eq:controller_storage_identity}. Applying the Cauchy--Schwarz inequality gives
\[
\|e_y\|^2 \le \left(\frac{b^2}{a}+g\right)(a\|\mathrm P_{\mathcal M_y^\perp}L\nabla V\|^2 + g\|\mathrm P_{\mathcal M_y^\perp}e_u\|^2),
\]
which, together with the triangle inequality $\|e_y\|\le b\|\mathrm P_{\mathcal M_y^\perp}L\nabla V\|+g\|\mathrm P_{\mathcal M_y^\perp}e_u\|$, establishes the inequality in~\eqref{eq:controller_storage_identity}. Finally, when $u\in\mathcal M_u$, we have $e_u=0$ and $\dot\xi = -a\,L^\top\mathrm P_{\mathcal M_y^\perp}L\nabla V$. Taking the inner product with $\nabla V$ gives $\langle\nabla V,\dot\xi\rangle=-a\|\mathrm P_{\mathcal M_y^\perp}L\nabla V\|^2$. Hence, $\dot\xi=0$ if and only if $\mathrm P_{\mathcal M_y^\perp}L\nabla V=0$.
\end{proof}

The projector $\mathrm P_{\mathcal M_y^\perp}$ in the state dynamics of~\eqref{eq:general_tp_controller} is essential: state directions with $L(u)\nabla V(\xi)\in\mathcal M_y$ are neither driven by $e_u$ nor damped, so the system can hold a nonzero state at equilibrium while its output remains on the target $\mathcal M_y$. In the formation control setting, these directions sustain the nonzero equilibrium edge forces in $\mathcal Z$.

\subsection{Edge Controller and Target Equilibria}
To address Problem~\ref{prob:synthesis}, we specialize Proposition~\ref{prop:controller_certificate} to the edge controller $\Sigma_2$ with port targets $(\mathcal M_{u_2},\mathcal M_{y_2})=(\mathcal C,\mathcal Z)$. To generate radial edge forces, we assign one scalar state $\xi_k\in\mathbb R$ to each edge $e_k\in\mathcal E$, stacked as $\xi\in\mathbb R^m$, with quadratic storage $V_2(\xi)=\frac12\|\xi\|^2$, and choose $L(u_2)=U(u_2)$ to align each $\xi_k$ with the unit edge direction $\widehat u_{2,k}$. With $e_{u_2}(u_2)=U(u_2)\ell$ from~\eqref{eq:sigma2_targetmaps}, the realization~\eqref{eq:general_tp_controller} yields the output-strictly $(\mathcal C,\mathcal Z)$-passive edge controller
\begin{equation}\label{eq:controller_synthesis}
\Sigma_2:\quad
\begin{cases}
\dot\xi = U(u_2)^\top \mathrm P_{\mathcal Z^\perp} U(u_2)\,\bigl(b\,\ell - a\,\xi\bigr),\\
y_2 = U(u_2)\,\bigl(b\,\xi + g\,\ell\bigr).
\end{cases}
\end{equation}
Componentwise, edge $e_k$ outputs the radial force $y_{2,k}=\widehat u_{2,k}(b\,\xi_k+g\,\ell_k)$, and its state evolves as
\begin{equation}\label{eq:edgewise_controller}
\dot\xi_k=\widehat u_{2,k}^\top\sum_{j=1}^m[\mathrm P_{\mathcal Z^\perp}]_{kj}\,\widehat u_{2,j}\,(b\,\ell_j-a\,\xi_j),
\end{equation}
where $[\mathrm P_{\mathcal Z^\perp}]_{kj}\in\mathbb R^{2\times2}$ denotes the $(k,j)$ block of $\mathrm P_{\mathcal Z^\perp}$. Hence, edge $e_k$ requires $\widehat u_{2,j}$, $\ell_j$, and $\xi_j$ only from the edges $e_j$ with $[\mathrm P_{\mathcal Z^\perp}]_{kj}\neq0$, so the block sparsity of $\mathrm P_{\mathcal Z^\perp}$ determines the inter-edge coupling pattern. For $\mathcal Z=\mathcal Z_0\otimes\mathbb R^2$ with $\mathcal Z_0\subseteq\mathbb R^m$, we have $\mathrm P_{\mathcal Z^\perp}=\mathrm P_{\mathcal Z_0^\perp}\otimes I_2$, and~\eqref{eq:edgewise_controller} reduces to $\dot\xi_k=\sum_{j}[\mathrm P_{\mathcal Z_0^\perp}]_{kj}\,(\widehat u_{2,k}^\top\widehat u_{2,j})(b\,\ell_j-a\,\xi_j)$, which involves only the relative orientations of the edges.

Interconnecting $\Sigma_1$ and $\Sigma_2$ via Fig.~\ref{fig:feedback_interconnection} (with $U$, $\ell$, and $e_{u_2}$ evaluated at $u_2=\bar E^\top p\in\mathcal N_{u_2}$) and defining the normalized rigidity matrix and controller Gram matrix
$$
R(p):= U(\bar E^\top p)^\top \bar E^\top, \;
G_{\mathcal Z}(p):= U(\bar E^\top p)^\top \mathrm P_{\mathcal Z^\perp}\,U(\bar E^\top p),
$$
we obtain the closed-loop dynamics
\begin{equation}\label{eq:closed_loop_dynamics}
\begin{cases}
\dot p = -R(p)^\top\bigl(b\,\xi + g\,\ell\bigr),\\
\dot\xi = G_{\mathcal Z}(p)\bigl(b\,\ell - a\,\xi\bigr).
\end{cases}
\end{equation}
Since $R_{\mathcal G}(p) = \diag\bigl(\|(\bar E^\top p)_1\|,\ldots,\|(\bar E^\top p)_m\|\bigr)R(p)$, infinitesimal rigidity implies $\ker R(p) = T_p\mathcal M_{y_1}$ and $\rank R(p) = 2n-3$ for all $p\in\mathcal M_{y_1}$.



We next characterize the closed-loop equilibria on $\mathcal M_{y_1}$ and the sustained equilibrium edge forces. 
For any $p$ with $\bar E^\top p\in\mathcal N_{u_2}$, since $\mathcal Z\subseteq\ker \bar E$ and $R(p)^\top = \bar E U(\bar E^\top p)$, we have
\begin{equation}\label{eq:radial_target_modes}
\thickmuskip=4mu\relax
\ker G_{\mathcal Z}(p) = \{\xi\in\mathbb R^m\colon U(\bar E^\top p)\xi\in\mathcal Z\} \subseteq \ker R(p)^\top.
\end{equation}
At any $p\in\mathcal M_{y_1}$ (so $\bar E^\top p\in\mathcal C$ and $\ell=0$), Proposition~\ref{prop:controller_certificate} implies that $\dot\xi=0$ if and only if $\xi\in\ker G_{\mathcal Z}(p)$, which by~\eqref{eq:radial_target_modes} also yields $\dot p = -b\,R(p)^\top\xi = 0$, with equilibrium edge forces $y_2=b\,U(\bar E^\top p)\xi\in\mathcal Z$. Since $\ker R(p)^\top = \diag(d)\ker(R_{\mathcal G}(p)^\top)$ on $\mathcal M_{y_1}$ is the normalized self-stress space, the controller equilibria $\ker G_{\mathcal Z}(p)$ at $p\in\mathcal M_{y_1}$ are precisely the normalized self-stresses whose induced edge forces lie in $\mathcal Z$, with $\ker G_{\ker \bar E}(p) = \ker R(p)^\top$. 

Moreover, for a held $p\in\mathcal M_{y_1}$, $\Sigma_2$ in~\eqref{eq:controller_synthesis} reduces to $\dot\xi = -a\,G_{\mathcal Z}(p)\xi$ with $e_{y_2} = b\,\mathrm P_{\mathcal Z^\perp}U(\bar E^\top p)\xi$. This shows the selective role of output-strict $(\mathcal C,\mathcal Z)$-passivity: it leaves the equilibrium modes in $\ker G_{\mathcal Z}(p)$ unpenalized while exponentially damping the orthogonal modes in $(\ker G_{\mathcal Z}(p))^\perp$ and the off-target output error $e_{y_2}$.

In general, $G_{\mathcal Z}(p)$ depends on $p$ through the edge directions. Rotating a configuration by $Q\in SO(2)$ maps $U(\bar E^\top p)$ to $(I_m\otimes Q)U(\bar E^\top p)$, so $G_{\mathcal Z}$ changes to $U(\bar E^\top p)^\top(I_m\otimes Q)^\top\mathrm P_{\mathcal Z^\perp}(I_m\otimes Q)U(\bar E^\top p)$, which coincides with $G_{\mathcal Z}(p)$ only if $\mathcal Z$ is rotation-invariant. Moreover, the convergence analysis requires the damped modes, i.e., the positive spectrum of $G_{\mathcal Z}(p)$, to remain uniformly damped near $\mathcal M_{y_1}$. We therefore impose the following assumption.

\begin{assumption}\label{ass:feasible_gap}
The subspace $\mathcal Z\subseteq\ker \bar E$ satisfies:
\begin{enumerate}
  \item[i)] \emph{Rotation invariance:} $(I_m\otimes Q)\mathcal Z = \mathcal Z$ for all $Q\in SO(2)$;
  \item[ii)] \emph{Uniform spectral gap:} there exist $\rho>0$ and $\mu_{\mathcal Z}\in(0,1]$ such that on $$\mathcal T_\rho:= \{p\in\mathbb R^{2n} : \dist(p,\mathcal M_{y_1})<\rho\}\subseteq\mathcal N_{y_1},$$ we have $\bar E^\top p\in\mathcal N_{u_2}$ and every positive eigenvalue of $G_{\mathcal Z}(p)$ lies in $[\mu_{\mathcal Z},1]$.
\end{enumerate}
\end{assumption}
Assumption~\ref{ass:feasible_gap}(i) holds for any $\mathcal Z = \mathcal Z_0\otimes\mathbb R^2$ with $\mathcal Z_0\subseteq\ker E$ (e.g., $\mathcal Z=\{0\}$ or $\ker \bar E = \ker E\otimes\mathbb R^2$). Under Assumption~\ref{ass:feasible_gap}(i), $\mathrm P_{\mathcal Z^\perp}$ commutes with $I_m\otimes Q$, so $G_{\mathcal Z}(p)$ is invariant under planar rigid motions; in particular, $G_{\mathcal Z}(p) = G_{\mathcal Z}(p^\star)$ on $\mathcal M_{y_1}$, and the closed-loop equilibrium set on $\mathcal M_{y_1}$ is
\begin{equation}\label{eq:target_equilibria}
\mathcal X_\star := \mathcal M_{y_1} \times \ker G_{\mathcal Z}(p^\star).
\end{equation}
Since $\bar E^\top\mathcal M_{y_1}\subseteq\mathcal C$ and $\mathcal Z\subseteq\ker \bar E$, $\mathcal X_\star$ also coincides with the closed-loop feasible state set in the sense of~\cite{yue2026Tpassivity}, on which all four realized ports lie in $(\mathcal M_{u_1},\mathcal M_{y_1},\mathcal C,\mathcal Z)$. Moreover, by this invariance, Assumption~\ref{ass:feasible_gap}(ii) holds (for $\rho$ sufficiently small) whenever $G_{\mathcal Z}(p)$ has constant rank near $p^\star$.

\subsection{Available Dissipation and the Realized Margin}
Along the closed-loop trajectories, the controller dissipation equality in~\eqref{eq:controller_storage_identity} reads $\dot V_2 = \langle e_{u_2},e_{y_2}\rangle - W_0(p,\xi)$, with
\begin{equation}\label{eq:exact_available_dissipation}
W_0(p,\xi) := a\,\xi^\top G_{\mathcal Z}(p)\xi + g\,\ell^\top G_{\mathcal Z}(p)\ell.
\end{equation}

Following~\cite{yue2026Tpassivity}, we retain this exact available dissipation $W_0(p,\xi)$ rather than the more conservative bound $\frac{a}{b^2+ag}\|e_{y_2}\|^2$ in~\eqref{eq:controller_storage_identity}. We next characterize the channel compatibility of the interconnection in Fig.~\ref{fig:feedback_interconnection}.


\begin{proposition}[Channel compatibility]\label{prop:channel_compatibility}
Consider the interconnection Fig.~\ref{fig:feedback_interconnection} with port targets~\eqref{eq:input_target_plant}, \eqref{eq:formation_target}, and~\eqref{eq:input_target}.
\begin{enumerate}
  \item[i)] The reverse channel $-\bar E$ is exactly compatible on $\mathbb R^{2m}$.
  \item[ii)] The forward channel $\bar E^\top$ has the defect
  \begin{equation}\label{eq:formation_defects}
  \begin{aligned}
  \delta_{\mathrm{fc}}(p) &= e_{u_2}(\bar E^\top p) - \bar E^\top e_{y_1}(p)\\
  &= \bar E^\top r_{y_1}(p) - r_{u_2}(\bar E^\top p),
  \end{aligned}
  \end{equation}
  which vanishes on $\mathcal M_{y_1}$. However, for $n\ge3$, the forward channel is not first-order compatible along any nonempty subset of $\mathcal M_{y_1}$, and hence not exactly compatible on any neighborhood of a point of $\mathcal M_{y_1}$.
  \item[iii)] For every $p\in\mathcal N_{y_1}$ with $\bar E^\top p\in\mathcal N_{u_2}$, the radial component of the forward defect is of second order:
  \begin{equation}\label{eq:realized_geometry}
  \bigl\|U(\bar E^\top p)^\top\delta_{\mathrm{fc}}(p)\bigr\|\le C\|e_{y_1}(p)\|^2,
  \end{equation}
  with $C:=2\|\bar E\|^2/\min_k d_k$.
\end{enumerate}
\end{proposition}

\begin{proof}
i) Since $\mathcal M_{u_1}=\{0\}$, we have $r_{u_1}\equiv0$, and $\mathcal Z\subseteq\ker\bar E$ gives $\delta_{\mathrm{rc}}(y_2)=-\bar E\,\mathrm P_{\mathcal Z}y_2-r_{u_1}(-\bar Ey_2)=0$ for all $y_2\in\mathbb R^{2m}$.

ii) For $p\in\mathcal M_{y_1}$, we have $r_{y_1}(p)=p$ and $\bar E^\top p\in\mathcal C$, so $r_{u_2}(\bar E^\top p)=\bar E^\top p$ and $\delta_{\mathrm{fc}}(p)=0$. At such $p$, the differentials of the nearest-point projections are the tangent-space projectors, so $\mathrm d_p\delta_{\mathrm{fc}}=\bar E^\top\mathrm P_{T_p\mathcal M_{y_1}}-\mathrm P_{T_{\bar E^\top p}\mathcal C}\,\bar E^\top$. Let $J$ denote the rotation by $\pi/2$. Since $T_p\mathcal M_{y_1}$ is spanned by the translations, which lie in $\ker\bar E^\top$, and the infinitesimal rotation $(I_n\otimes J)p$, with $\bar E^\top(I_n\otimes J)p=(I_m\otimes J)\bar E^\top p\neq0$, the first term has rank one. For the second term, $\mathrm P_{T_{\bar E^\top p}\mathcal C}=I_{2m}-UU^\top$ with $U=U(\bar E^\top p)$, so $v\in\ker(\mathrm P_{T_{\bar E^\top p}\mathcal C}\bar E^\top)$ if and only if each relative velocity $(\bar E^\top v)_k$ is parallel to $(\bar E^\top p)_k$, i.e., $R_{\mathcal G}(p)(I_n\otimes J)^\top v=0$. By infinitesimal rigidity, this kernel is three-dimensional, so the second term has rank $2n-3\ge3$. Hence, $\mathrm d_p\delta_{\mathrm{fc}}\neq0$ for every $p\in\mathcal M_{y_1}$, and $\delta_{\mathrm{fc}}$ cannot vanish identically near any such $p$.

iii) Fix an edge $e_k$ and let $\bar u_k:=(\bar E^\top r_{y_1}(p))_k$ denote the corresponding edge of the nearest target formation. Since $r_{y_1}(p)\in\mathcal M_{y_1}$, both $\bar u_k$ and $r_{u_2,k}=d_k\widehat u_{2,k}$ lie on the circle $d_kS^1$, and $\delta_{\mathrm{fc},k}=\bar u_k-d_k\widehat u_{2,k}$. Expanding the squared norm with $\|\bar u_k\|=d_k$ gives
\[
\begin{aligned}
\|\delta_{\mathrm{fc},k}\|^2
&= \|\bar u_k\|^2-2d_k\widehat u_{2,k}^\top\bar u_k+d_k^2\\
&= 2d_k\bigl(d_k-\widehat u_{2,k}^\top\bar u_k\bigr)
= -2d_k\,\widehat u_{2,k}^\top\delta_{\mathrm{fc},k},
\end{aligned}
\]
so the radial component of each chord is quadratic in its length,
\[
\bigl|\widehat u_{2,k}^\top\delta_{\mathrm{fc},k}\bigr|=\frac{\|\delta_{\mathrm{fc},k}\|^2}{2d_k}.
\]
Next, since $\bar E^\top r_{y_1}(p)\in\bar E^\top\mathcal M_{y_1}\subseteq\mathcal C$, the nearest-point property of $r_{u_2}$ gives
\[
\begin{aligned}
\|e_{u_2}(\bar E^\top p)\|&=\|\ell\|=\dist(\bar E^\top p,\mathcal C)\\
&\le\|\bar E^\top e_{y_1}(p)\|\le\|\bar E\|\,\|e_{y_1}(p)\|,
\end{aligned}
\]
and hence, by~\eqref{eq:formation_defects},
\[
\|\delta_{\mathrm{fc}}\|\le\|e_{u_2}\|+\|\bar E^\top e_{y_1}\|\le2\|\bar E\|\,\|e_{y_1}\|.
\]
Combining these bounds yields
\[
\begin{aligned}
\|U^\top\delta_{\mathrm{fc}}\|
&\le\sum_{k=1}^m\bigl|\widehat u_{2,k}^\top\delta_{\mathrm{fc},k}\bigr|
=\sum_{k=1}^m\frac{\|\delta_{\mathrm{fc},k}\|^2}{2d_k}\\
&\le\frac{\|\delta_{\mathrm{fc}}\|^2}{2\min_kd_k}
\le C\|e_{y_1}\|^2.
\end{aligned}
\]
\end{proof}

By Proposition~\ref{prop:channel_compatibility}ii), the interconnection results of~\cite{yue2026Tpassivity} for exactly or first-order compatible channels do not apply. Intuitively, the edgewise target $\mathcal C$ allows each edge to rotate independently on its own circle, whereas on $\bar E^\top\mathcal M_{y_1}$ all edges can only rotate together.  Therefore, off $\mathcal M_{y_1}$, the two projections in~\eqref{eq:formation_defects} differ to first order, but only along the edge-rotation (tangential) directions. Item~iii) is the structural property that nevertheless yields a dissipation margin: although $\delta_{\mathrm{fc}}$ is of first order in general, the controller output~\eqref{eq:controller_synthesis} is radial, so $\delta_{\mathrm{fc}}$ enters the energy balance only through its second-order radial component (cf.\ Fig.~\ref{fig:edge_projection}).

Since $\delta_{\mathrm{rc}}\equiv 0$ by Proposition~\ref{prop:channel_compatibility} i), the composite storage $V_{\mathrm{cl}}(p,\xi) = V_1(p) + V_2(\xi) = \frac12\|e_{y_1}(p)\|^2 + \frac12\|\xi\|^2$ satisfies the closed-loop dissipation identity
\begin{equation}\label{eq:closed_loop_dissipation}
\dot V_{\mathrm{cl}} = \langle\delta_{\mathrm{fc}}(p), e_{y_2}\rangle - W_0(p,\xi) =: s_\delta(p,\xi) - W_0(p,\xi).
\end{equation}
Comparing $s_\delta$ with $W_0$ yields the following margin.

\begin{lemma}[Dissipation margin]\label{lem:net_dissipation}
Under Assumption~\ref{ass:feasible_gap} with spectral bound $\mu_{\mathcal Z}$, if the controller gains satisfy
\begin{equation}\label{eq:gain_certificate}
4ag\mu_{\mathcal Z} > b^2(1-\mu_{\mathcal Z})^2,
\end{equation}
then there exist $r\in(0,\rho]$ and $\alpha>0$ such that, for every $p\in\mathcal T_r$ and $\xi\in\mathbb R^m$,
\begin{equation}\label{eq:confined_dissipation}
\dot V_{\mathrm{cl}}
\le -\alpha\bigl(\|e_{y_1}(p)\|^2 + \xi^\top G_{\mathcal Z}(p)\xi\bigr).
\end{equation}
\end{lemma}

\begin{proof}
For brevity, write $U=U(\bar E^\top p)$, $G=G_{\mathcal Z}(p)$, $R=R(p)$, $e=e_{y_1}(p)$, and $\delta=\delta_{\mathrm{fc}}(p)$, and define
\[
X:=\sqrt{\xi^\top G\xi},\; Y:=\|Re\|.
\]

\emph{Step~1 (energy balance).} Since $e_{u_2}=U\ell$, we have $\delta=U\ell-\bar E^\top e$, and $U^\top U=I_m$ with $R=U^\top\bar E^\top$ gives
\[
\ell=Re+U^\top\delta.
\]
Moreover, $e_{y_2}=\mathrm P_{\mathcal Z^\perp}U(b\xi+g\ell)$ and $\bar E\,\mathrm P_{\mathcal Z}=0$ give
\[
\begin{aligned}
\langle U\ell,e_{y_2}\rangle&=\ell^\top G(b\xi+g\ell)\\
\quad\langle\bar E^\top e,e_{y_2}\rangle&=(Re)^\top(b\xi+g\ell).
\end{aligned}
\]
Substituting these relations into~\eqref{eq:closed_loop_dissipation} yields
\begin{equation}\label{eq:realized_storage_identity}
\begin{aligned}
\dot V_{\mathrm{cl}}
&= -a\,\xi^\top G\xi - g\|Re\|^2 - b(Re)^\top(I_m - G)\xi \\
&\quad{} - (U^\top\delta)^\top\bigl(g\,Re - b\,G\xi\bigr).
\end{aligned}
\end{equation}

\emph{Step~2 (quadratic terms).} By~\eqref{eq:radial_target_modes}, every $\zeta\in\ker G$ satisfies $U\zeta\in\mathcal Z\subseteq\ker\bar E$, so $\zeta^\top Re=(\bar EU\zeta)^\top e=0$, i.e., $Re\perp\ker G$. Since the positive eigenvalues of $G$ lie in $[\mu_{\mathcal Z},1]$ by Assumption~\ref{ass:feasible_gap}(ii), it follows that
\[
\bigl|(Re)^\top(I_m-G)\xi\bigr|\le\frac{1-\mu_{\mathcal Z}}{\sqrt{\mu_{\mathcal Z}}}\,XY.
\]
Hence, the first three terms in~\eqref{eq:realized_storage_identity} are bounded by the quadratic form $-aX^2-gY^2+b\frac{1-\mu_{\mathcal Z}}{\sqrt{\mu_{\mathcal Z}}}XY$, which is negative definite under~\eqref{eq:gain_certificate}. Thus, there exists $\alpha_0>0$ such that
\[
-a X^2-gY^2-b(Re)^\top(I_m-G)\xi\le-2\alpha_0(X^2+Y^2).
\]

\emph{Step~3 (lower bound on $Y$).} Since $e\in N_{r_{y_1}(p)}\mathcal M_{y_1}=(\ker R(r_{y_1}(p)))^\perp$, infinitesimal rigidity and the invariance of $\mathcal M_{y_1}$ under planar rigid motions give a constant $\sigma_\star>0$ such that $\|R(r_{y_1}(p))e\|\ge\sigma_\star\|e\|$. Since $R(p)e=R(r_{y_1}(p))e+O(\|e\|^2)$, there exist $r\in(0,\rho]$ and $c_1>0$ such that
\[
Y\ge c_1\|e\|\qquad\text{for all }p\in\mathcal T_r.
\]

\emph{Step~4 (remainder).} By Proposition~\ref{prop:channel_compatibility}iii), $\|U^\top\delta\|\le C\|e\|^2$. Together with $\|G\xi\|\le X$, which follows from $0\preceq G\preceq I_m$, and Step~3, the last term in~\eqref{eq:realized_storage_identity} satisfies
\[
\begin{aligned}
\bigl|(U^\top\delta)^\top(g\,Re-b\,G\xi)\bigr|
&\le C\|e\|^2\,(gY+bX)\\
&\le C'\|e\|\,\bigl(X^2+Y^2\bigr),
\end{aligned}
\]
with $C':=C(2g+b)/(2c_1)$. Shrinking $r$ so that $C'r\le\alpha_0$ and combining Steps~1--4 yields $\dot V_{\mathrm{cl}}\le-\alpha_0(X^2+Y^2)$ on $\mathcal T_r$. By Step~3, this implies~\eqref{eq:confined_dissipation} with $\alpha:=\alpha_0\min\{1,c_1^2\}$.
\end{proof}

\begin{remark}[Role of the gain condition]\label{rem:gain_condition}
For $\mathcal Z\neq\{0\}$, the residual supply $s_\delta$ in~\eqref{eq:closed_loop_dissipation} retains a nonzero quadratic part near $\mathcal X_\star$. Indeed, at any $(\bar p,0)$ with $\bar p\in\mathcal M_{y_1}$, writing $\bar R:=R(\bar p)$ and $\bar G:=G_{\mathcal Z}(\bar p)$, a state perturbation $(v,\eta)$ yields
\[
s_\delta(\bar p+tv,\,t\eta) = -t^2(\bar Rv)^\top(I_m-\bar G)(b\,\eta+g\bar Rv) + O(t^3),
\]
whose quadratic part is nonzero whenever $\bar R^\top(I_m-\bar G)\ne 0$. The closed loop thus falls into the \emph{nonzero quadratic residual} case of the compatibility hierarchy in~\cite{yue2026Tpassivity}, so part of the available dissipation $W_0$ must be spent to dominate $s_\delta$; this is the origin of the gain condition~\eqref{eq:gain_certificate}. For $\mathcal Z=\{0\}$, we have $\bar G=I_m$ and the quadratic part cancels (\emph{quadratic cancellation}), consistent with Corollary~\ref{cor:zero_output_convergence} holding for all gains $a,b,g>0$.
\end{remark}

\subsection{Compact Confinement and Convergence}
Fixing $r\in(0,\rho]$ from Lemma~\ref{lem:net_dissipation}, we define the \emph{closed-loop analysis domain} $\mathcal D:=\mathcal T_r\times\mathbb R^m$, on which~\eqref{eq:confined_dissipation} holds. However, Lemma~\ref{lem:net_dissipation} alone does not guarantee convergence:~\eqref{eq:confined_dissipation} holds only locally and leaves planar translations along $\mathcal M_{y_1}$ and modes in $\ker G_{\mathcal Z}(p)$ unpenalized, rendering the sublevel sets of $V_{\mathrm{cl}}$ noncompact. We next combine these sublevel sets with the conserved formation centroid to obtain compact confinement and closed-loop convergence.

\begin{theorem}[Local state-set and port convergence]\label{thm:local_convergence}
Consider the closed loop of $\Sigma_1$ and $\Sigma_2$ interconnected via Fig.~\ref{fig:feedback_interconnection}. Suppose Assumption~\ref{ass:feasible_gap} and the gain condition~\eqref{eq:gain_certificate} hold. Then, there exists $\varepsilon_0>0$ such that every solution with $(p(0),\xi(0))\in\mathcal D$ and $V_{\mathrm{cl}}(p(0),\xi(0))<\varepsilon_0$ is forward complete, remains in a compact subset of $\mathcal D$, and satisfies
\begin{equation}\label{eq:state_set_convergence}
\lim_{t\to\infty}\dist\bigl((p(t),\xi(t)),\mathcal X_\star\bigr) = 0.
\end{equation}
Moreover, the port signals $u_1$, $y_1$, $u_2$, and $y_2$ converge to $\mathcal M_{u_1}$, $\mathcal M_{y_1}$, $\mathcal C$, and $\mathcal Z$, respectively.
\end{theorem}

\begin{proof}
Choose $\varepsilon_0\in(0,r^2/2)$ and let $c:=V_{\mathrm{cl}}(p(0),\xi(0))<\varepsilon_0$. Since $(\mathbf1_n^\top\otimes I_2)\bar E=0$, the centroid $\chi(p):=(\mathbf1_n^\top\otimes I_2)p$, with $\dot\chi=0$, is conserved. On $\{V_{\mathrm{cl}}\le c\}$, we have $\dist(p,\mathcal M_{y_1})=\|e_{y_1}(p)\|\le\sqrt{2c}<r$ and $\|\xi\|\le\sqrt{2c}$. Since $\mathcal M_{y_1}$ is translation-invariant, $r_{y_1}(p)$ preserves $\chi$, so fixing $\chi(p)=\chi(p(0))$ restricts $r_{y_1}(p)$ to a compact $SO(2)$-orbit. Thus, $\Omega_c:=\{(p,\xi)\in\mathcal D: V_{\mathrm{cl}}(p,\xi)\le c,\ \chi(p)=\chi(p(0))\}$ is a compact subset of $\mathcal D$. Moreover, $\Omega_c$ is forward invariant: as long as $p(t)\in\mathcal T_r$,~\eqref{eq:confined_dissipation} gives $\dot V_{\mathrm{cl}}\le0$, so $\dist(p(t),\mathcal M_{y_1})\le\sqrt{2c}<r$ and the trajectory cannot reach the boundary of $\mathcal T_r$. This ensures forward completeness and compact confinement by~\cite[Proposition~6]{yue2026Tpassivity}.
Next, since $\|e_{u_2}(\bar E^\top p)\|=\|\ell\|\le\|\bar E\|\|e_{y_1}(p)\|$ from the proof of Proposition~\ref{prop:channel_compatibility} iii), it follows from~\eqref{eq:controller_synthesis} that
\begin{equation}\label{eq:port_error_bounds}
\|e_{y_2}(y_2)\| \le b\sqrt{\xi^\top G_{\mathcal Z}(p)\xi} + g\|\ell\|,\ \|u_1\| \le \|\bar E\|\|e_{y_2}(y_2)\|,
\end{equation}
and, since the squared distance to a product set is the sum of the squared distances to its factors, $\dist\bigl((p,y_2),\mathcal M_{y_1}\times\mathcal Z\bigr)^2 = \|e_{y_1}(p)\|^2+\|e_{y_2}(y_2)\|^2 \le (1+2g^2\|\bar E\|^2)\|e_{y_1}(p)\|^2 + 2b^2\xi^\top G_{\mathcal Z}(p)\xi$. Hence,~\eqref{eq:confined_dissipation} satisfies the port-distance margin of~\cite[Theorem~2]{yue2026Tpassivity}, yielding $(p,y_2)\to\mathcal M_{y_1}\times\mathcal Z$ and, by~\eqref{eq:port_error_bounds}, $(u_1,u_2)\to\mathcal M_{u_1}\times\mathcal C$. Finally, $\ell\to0$ and $e_{y_2}=\mathrm P_{\mathcal Z^\perp}U(b\xi+g\ell)\to0$ imply $\xi^\top G_{\mathcal Z}(p)\xi=\|\mathrm P_{\mathcal Z^\perp}U\xi\|^2\to0$. Hence, every accumulation point $(\bar p,\bar\xi)\in\Omega_c$ of the trajectory satisfies $\bar p\in\mathcal M_{y_1}$ and $\bar\xi\in\ker G_{\mathcal Z}(\bar p)=\ker G_{\mathcal Z}(p^\star)$, i.e., lies in $\mathcal X_\star$ by~\eqref{eq:target_equilibria}, and compactness of $\Omega_c$ yields~\eqref{eq:state_set_convergence}.
\end{proof}

We next specialize Theorem~\ref{thm:local_convergence} to $\mathcal Z=\{0\}$. In this case, $\mathrm P_{\mathcal Z^\perp}=I_{2m}$, so that the controller~\eqref{eq:controller_synthesis} decouples across edges into $\dot\xi_k = b\ell_k - a\xi_k$ and $y_{2,k} = \widehat u_{2,k}(b\xi_k + g\ell_k)$ for $k\in\{1,\ldots,m\}$, and the target equilibrium set reduces to $\mathcal X_\star=\mathcal M_{y_1}\times\{0\}$.

\begin{corollary}[Independent edge controllers]\label{cor:zero_output_convergence}
Consider the closed loop of $\Sigma_1$ and $\Sigma_2$ interconnected via Fig.~\ref{fig:feedback_interconnection} with $\mathcal Z=\{0\}$ and any $a,b,g>0$. Then there exists $\varepsilon_0>0$ such that every solution with $(p(0),\xi(0))\in\mathcal D$ and $V_{\mathrm{cl}}(p(0),\xi(0))<\varepsilon_0$ converges exponentially to $\mathcal M_{y_1}\times\{0\}$, and $p(t)$ converges to a single configuration in $\mathcal M_{y_1}$.
\end{corollary}

\begin{proof}
When $\mathcal Z=\{0\}$, we have $G_{\mathcal Z}(p)=I_m$ and $\mu_{\mathcal Z}=1$, so Assumption~\ref{ass:feasible_gap} and~\eqref{eq:gain_certificate} hold for all $a,b,g>0$. By Theorem~\ref{thm:local_convergence}, for $\varepsilon_0\in(0,r^2/2)$, every solution with $V_{\mathrm{cl}}(0)<\varepsilon_0$ remains in $\mathcal D$ for all $t\ge 0$, where~\eqref{eq:confined_dissipation} reduces to $\dot V_{\mathrm{cl}}\le -2\alpha V_{\mathrm{cl}}$. Consequently, $V_{\mathrm{cl}}(t)\le V_{\mathrm{cl}}(0)e^{-2\alpha t}$ and
$\dist\bigl((p(t),\xi(t)),\mathcal M_{y_1}\times\{0\}\bigr) = \sqrt{2V_{\mathrm{cl}}(t)} \le \sqrt{2V_{\mathrm{cl}}(0)}\,e^{-\alpha t}$,
establishing exponential convergence to $\mathcal M_{y_1}\times\{0\}$. Moreover, by~\eqref{eq:port_error_bounds}, $\|\dot p(t)\|=\|u_1(t)\|\le C\sqrt{2V_{\mathrm{cl}}(t)}\le C\sqrt{2V_{\mathrm{cl}}(0)}\,e^{-\alpha t}$ for some $C>0$, which is integrable on $[0,\infty)$. Thus, $p(t)$ converges to a single configuration in $\mathcal M_{y_1}$.
\end{proof}

While $\mathcal Z=\{0\}$ in Corollary~\ref{cor:zero_output_convergence} eliminates inter-edge coupling with $\mu_{\{0\}}=1$, it sustains no nonzero equilibrium forces. At the other extreme, $\mathcal Z=\ker \bar E$ retains the full self-stress space $\ker R(p^\star)^\top$, but its projector $\mathrm P_{(\ker \bar E)^\perp}$ is generally dense and couples nonadjacent edges. Whenever a smaller subspace $\mathcal Z\subsetneq\ker\bar E$ has the same intersection with $\range U(\bar E^\top p)$ as $\ker\bar E$ near $\mathcal M_{y_1}$, we have $G_{\mathcal Z}(p)\succeq G_{\ker\bar E}(p)$ with the same kernel $\ker R(p)^\top$, which preserves the admissible equilibrium forces and admits $\mu_{\mathcal Z}\ge\mu_{\ker\bar E}$. In particular, when the desired equilibrium forces are supported on a local subgraph, choosing $\mathcal Z$ as the subspace of balanced forces supported on that subgraph confines the coupling of $\mathrm P_{\mathcal Z^\perp}$ to the subgraph and decouples the remaining edges, as illustrated in Section~\ref{sec:numerical_results}.

\section{Numerical results}\label{sec:numerical_results}
We illustrate three aspects of the results: how the output target $\mathcal Z$ selects equilibrium edge forces sustained at steady state, the second-order scaling of the forward compatibility defect in Proposition~\ref{prop:channel_compatibility}, and the role of the spectral bound $\mu_{\mathcal Z}$ in the convergence rate.

\begin{figure*}[t]
\centering
\begin{minipage}[c]{0.47\textwidth}
\centering
\includegraphics[width=\linewidth]{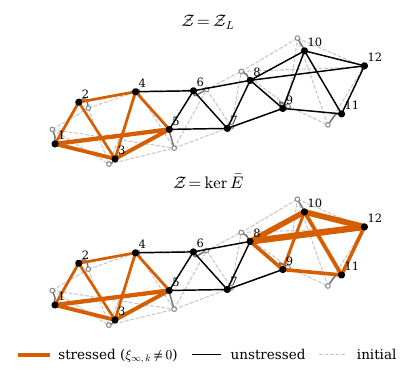}\\[-1pt]
{\footnotesize (a)}
\end{minipage}\hfill
\begin{minipage}[c]{0.5\textwidth}
\centering\footnotesize
\setlength{\tabcolsep}{3.5pt}
\begin{tabular}{@{}lccccc@{}}
\toprule
$\mathcal Z$ & stressed & coupled pairs & $\mu_{\mathcal Z}$ & local rate & $t_s$ (med./max) \\
\midrule
$\{0\}$ & 0 & 0 & 1.000 & 0.877 & 5.5 / 7.1 \\
$\mathcal Z_L$ & 8 & 45 & 0.131 & 0.136 & 21.6 / 28.0 \\
$\mathcal Z_R$ & 8 & 45 & 0.078 & 0.082 & 35.9 / 49.9 \\
$\mathcal Z_L\oplus\mathcal Z_R$ & 16 & 90 & 0.078 & 0.082 & 36.2 / 49.9 \\
$\ker\bar E$ & 16 & 253 & 0.011 & 0.012 & 248.9 / 368.9 \\
\bottomrule
\end{tabular}\\[3pt]
(b)
\caption{(a) Agent trajectories (gray) from the initial (dashed) to the final formation for $\mathcal Z=\mathcal Z_L$ (top) and $\mathcal Z=\ker\bar E$ (bottom); highlighted edges carry a nonzero sustained stress $\xi_{\infty,k}$, with line width proportional to $|\xi_{\infty,k}|$. (b) For each output target: number of stressed edges, edge pairs coupled by $\mathrm P_{\mathcal Z^\perp}$, spectral bound $\mu_{\mathcal Z}$ at $p^\star$, slowest local decay rate at $(p^\star,0)$, and settling time $t_s$ of $\|\ell(t)\|\le10^{-3}$ (median and maximum over $20$ random initial conditions), for $a=b=1$ and $g=25$.}
\label{fig:ex_formation}
\end{minipage}
\end{figure*}

Consider $n=12$ agents with target positions $p_i^\star=(0.9(i-1),\,y_i)$, where $(y_1,\ldots,y_{12})=(0,1.2,-0.7,1.3,0,1.1,-0.2,1.2,0.2,1.9,-0.2,1.2)$. The edges $\{i,i+1\}$ and $\{i,i+2\}$ form a chain of triangles, to which we add the two long-range links $\{1,5\}$ and $\{8,12\}$, giving $m=23$ edges (Fig.~\ref{fig:ex_formation}(a)). The framework is infinitesimally rigid, and each long-range link creates one independent self-stress: one is nonzero exactly on the eight edges among agents $1$--$5$, the other on the eight edges among agents $8$--$12$, and the seven edges in between carry no self-stress.

We split the agents into a left group $L=\{1,\ldots,6\}$ and a right group $R=\{7,\ldots,12\}$ and compare five output targets:
\begin{itemize}
\item $\mathcal Z=\{0\}$: no nonzero equilibrium force is retained;
\item $\mathcal Z_L$: the balanced edge forces in $\ker\bar E$ that vanish on every edge not joining two agents of $L$;
\item $\mathcal Z_R$: defined analogously for $R$;
\item $\mathcal Z_L\oplus\mathcal Z_R$: the sum of the two;
\item $\mathcal Z=\ker\bar E$: all balanced edge forces.
\end{itemize}
Each target has the form $\mathcal Z_0\otimes\mathbb R^2$ and thus satisfies Assumption~\ref{ass:feasible_gap}(i), and $\dim\ker G_{\mathcal Z}(p(t))$ stays constant along all simulated trajectories, consistent with Assumption~\ref{ass:feasible_gap}(ii). We use $a=b=1$ and $g=25$, which satisfy~\eqref{eq:gain_certificate} at $\mu_{\mathcal Z}(p^\star)$ for every target. Each target is simulated from the initial condition of Fig.~\ref{fig:ex_formation}(a) and from $20$ random ones, in which $p(0)$ is a randomly rotated and translated copy of $p^\star$ perturbed by Gaussian noise with standard deviation $0.25$, and $\xi(0)$ is random.

\emph{Sustained equilibrium forces.} In all runs, the formation converges to the target shape and $\xi(t)$ converges to a normalized self-stress $\xi_\infty\in\ker G_{\mathcal Z}(p^\star)$, inducing edge forces that balance at every agent. The output target determines which edges carry these forces (Fig.~\ref{fig:ex_formation}(a)): under $\mathcal Z_L$, only the eight edges among agents $1$--$5$ remain stressed, whereas under $\mathcal Z_L\oplus\mathcal Z_R$ and $\ker\bar E$, both groups of eight edges are stressed. Which edges are stressed is the same in every run of a given design; only the magnitude of $\xi_\infty$ depends on the initial condition. The stressed edges also have a simple interpretation: an edge can carry a nonzero equilibrium force exactly when it is redundant, i.e., when removing it preserves infinitesimal rigidity~\cite{asimow1978rigidity,connelly2022frameworks}. For example, the links $\{4,6\}$ and $\{5,6\}$ remain unstressed under $\mathcal Z_L$, and removing either of them makes the formation flexible.

\begin{figure*}[t]
\centering
\includegraphics[width=\textwidth]{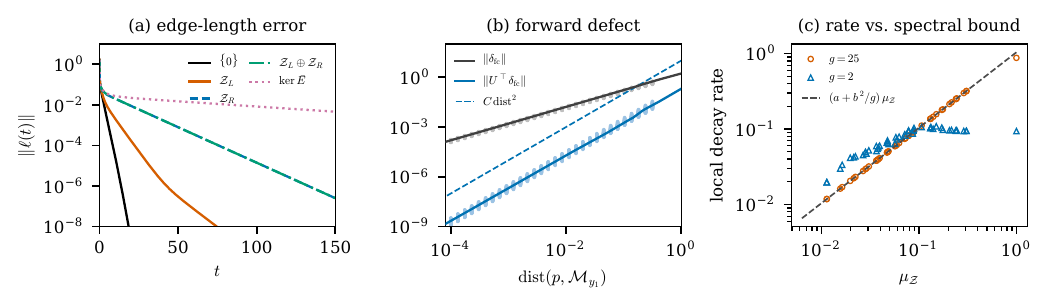}
\caption{(a) Edge-length error $\|\ell(t)\|=\|e_{u_2}(u_2(t))\|$ from the initial condition of Fig.~\ref{fig:ex_formation}(a); the curves for $\mathcal Z_R$ and $\mathcal Z_L\oplus\mathcal Z_R$ nearly coincide. (b) Forward compatibility defect versus $\dist(p,\mathcal M_{y_1})$ at sampled configurations (dots) and along the closed-loop trajectory for $\mathcal Z_L$ (solid). (c) Slowest local decay rate versus $\mu_{\mathcal Z}$ for $50$ output targets.}
\label{fig:ex_analysis}
\end{figure*}

\emph{Forward compatibility defect.} Fig.~\ref{fig:ex_analysis}(b) checks Proposition~\ref{prop:channel_compatibility} at configurations displaced from $p^\star$ in $40$ random directions normal to $\mathcal M_{y_1}$, and along the closed-loop trajectory for $\mathcal Z_L$. The full defect $\delta_{\mathrm{fc}}$ grows linearly with the distance to the target (fitted slope $1.00$), so the forward channel is not first-order compatible. Its radial component $U^\top\delta_{\mathrm{fc}}$, which is the only part that enters the closed-loop energy balance~\eqref{eq:realized_storage_identity}, grows quadratically (fitted slope $2.00$) and remains below the bound~\eqref{eq:realized_geometry} with $C=9.94$.

\emph{Spectral bound and convergence rate.} Fig.~\ref{fig:ex_formation}(b) and Fig.~\ref{fig:ex_analysis}(a) show that convergence slows down as $\mu_{\mathcal Z}$ decreases. To test this relation beyond the five designs, Fig.~\ref{fig:ex_analysis}(c) plots the slowest decay rate of the closed loop, linearized at $(p^\star,0)$, against $\mu_{\mathcal Z}$ for $50$ output targets: the five above, and the targets obtained in the same way as $\mathcal Z_L$ from every group of four or more consecutive agents. For $g=25$, the rate matches $(a+b^2/g)\,\mu_{\mathcal Z}$ within $0.5\%$ for every target except $\mathcal Z=\{0\}$: the agents settle quickly, and the controller states then decay at a rate set by $\mu_{\mathcal Z}$. For $g=2$, the rate follows $\mu_{\mathcal Z}$ until the agent dynamics become the bottleneck. 
The output target thus trades the sustained equilibrium forces against coupling and speed: $\mathcal Z_L\oplus\mathcal Z_R$ and $\ker\bar E$ sustain the same equilibrium forces, but $\ker\bar E$ couples all $253$ edge pairs instead of $90$, has a seven times smaller $\mu_{\mathcal Z}$, and settles about seven times more slowly.

\section{Concluding remarks}\label{sec:conclusion}
We develop a passivity-based dynamical synthesis for distance-constrained formations using transversal passivity. By encoding the desired formation geometry and balanced edge forces as port target manifolds, the control task is reformulated as the joint design of a transversally passive edge controller and its output-target subspace. We propose a constructive realization of an output-strictly transversally passive edge controller that sustains the corresponding equilibrium edge forces, and establish a closed-loop dissipation margin and local convergence to the target manifolds. While sustaining nonzero equilibrium forces generally leads to a centralized implementation, numerical simulations illustrate how a localized output-target subspace can preserve these forces with sparser coupling. Future work will explore sparse output-target designs for distributed controller implementation and develop a general synthesis framework for transversal passivity.
\bibliographystyle{IEEEtran}
\bibliography{ref}

\end{document}